\documentclass[
aps,%
12pt,%
final,%
notitlepage,%
oneside,%
onecolumn,%
nobibnotes,%
nofootinbib,%
superscriptaddress,%
noshowpacs,%
centertags]%
{revtex4}

\begin{document}

\title{
BETA- AND GAMMA-DECAY TRANSITION RATES IN THE
TWO-GROUP SHELL MODEL}

\author{{\firstname{V. I.} \surname{Isakov}}}
\email{Visakov@thd.pnpi.spb.ru}
\affiliation{
National Research Center\,\,``Kurchatov Institute'', \\
Petersburg Nuclear Physics Institute, Gatchina, Russia
}

\begin{abstract} \noindent
In the framework of the two-group configuration model we obtain formulas
for the reduced transition rates for beta- and gamma-transitions in
even--even, odd--odd, even--odd, and odd--even nuclei. We explored dependencies
of the transition rates on the occupancies of the involved subshells, as
well as on the spin values of the initial and final states. The obtained
formulas are useful for the qualitative spectroscopic analysis of experimental
data, particulary in the regions of magicity, including the regions of the
``remote" nuclei.
\end{abstract}
\maketitle

Study of beta- and gamma-transitions is one of the powerful methods of
nuclear spectroscopy which enables us to define quantum characteristics
of nuclear states implicated in the decay. The magnitude of the
corresponding reduced transition rate $B(\lambda)$ depends on the
multipolarity of the transition, as well as on the structure
of the wave functions of the initial and final states. This structure in
turn depends on the number of active neutrons and protons, on the
effective interaction between them, and on the values $I^{\pi}$ of
the involved states. We have the most definite knowledge on the structure
of states for nuclei that are not very far from the closed shells, where
at a first approximation it is described in terms of the multi-particle
shell model with no configuration mixing. This approach offers a good
reference point for a precise calculations, see \cite{Isakov09}, where
calculations for some nuclei just close to $N = Z$
were performed in the representation of the total isospin, and is useful for
qualitative evaluations. The corresponding formulas give us an opportunity
to reveal the dependence of the reduced transition rates on the occupancies
of single-particle levels and their single-particle characteristics, as
well as on the values of the total angular momenta of the involved nuclear
states.

In case of beta-decay of spherical nuclei that are considered here, two
$nljt_{z}$-orbitals are involved in the transition. For example,
in $\beta^{+}$-decay we have $j_{1}(p) \to j_{2}(n)$. In the general case,
we have the transformation between the configurations
\begin{eqnarray}
|i\rangle &=& \left|j_1^{n_1}(s_1\alpha_1J_1),j_2^{n_2}
(s_2\alpha_2J_2);I_i\right\rangle_a \,\,\, {\rm and}
\nonumber\\
|f\rangle &=& \left|
j_1^{n_1\!-\!1}(s'_1\alpha_1'J'_1), j^{n_2\!+\!1}_2
(s'_2\alpha'_2J'_2); I_f\right\rangle_a.
\end{eqnarray}

Here, we used the $jj$-coupling scheme and the neutron--proton
representation, where $j_{1}(p) \equiv n_{1},l_{1},j_{1},t_{z_{1}}= -1/2$, while
$j_{2}(n) \equiv n_{2},l_{2},j_{2},t_{z_{2}}=+1/2$.
In (1) $``s"$ is seniority (number of unpaired particles), while $``\alpha"$ is
an additional quantum number (if it is necessary for unambiguous classification
of state). The proposed structure of the
wave functions is a very good starting point in the case of short-range
attraction between the identical nucleons. Wave functions $|i\rangle$ and
$|f\rangle$ are antisymmetric with respect to all permutations, including
those between particles of the $\{ j_{1} \}$ and $\{ j_{2} \}$
groups.

To obtain formulas for reduced transition rates we use the formalism
of the fractional parentage decompositions. The corresponding algebra in
the case of one-group configuration is represented in \cite{Talmi}, while the
necessary formalism for the two-group configurations one can find in
\cite{Neudat, Nemetz, Hong-zhou} (in \cite{Neudat} for
the case of $LS$-coupling).

By making the necessary
recouplings and permutations, we can represent the functions $|i\rangle$ and
$|f\rangle$ in the form, that is suitable for calculation of matrix elements
of the single-particle transition operators:
\begin{eqnarray}
&& \hspace*{-0.5cm} |i\rangle=
\left| j^{n_1}_1\,(s_1\alpha_1J_1),\,j^{n_2}_2\, (s_2\alpha_2J_2);
I_i\,\right\rangle_a\ =
\\
&& =\sum_{s_3\alpha_3J_3J'}\!(-1)^{n_2+j_1\!+\!J_2\!+\!J_1\!+\!J'}
\sqrt{\frac{n_1(2J_1+1)(2J'+1)}{(n_1+n_2)}} \Big\{ {J_3\,j_1\,J_1 \atop
I_i\,J_2\,J'} \Big\}\ \times
\nonumber
\\
&&\times\Big[j_1^{n_1\!-\!1}(s_3\alpha_3J_3)j_1J_1|\}
j_1^{n_1}(s_1\alpha_1J_1)\Big] \left|
\left(j_1^{n_1\!-\!1}(s_3\alpha_3J_3),
j_2^{n_2}(s_2\alpha_2J_2)\right)_aJ',j_1;I_i\right\rangle +
\nonumber
\\
&&+ \!\!\!\!\!\sum_{s_4\alpha_4J_4J}\!\!\!(\!-\!1\!)^{J_1\!\!
+\!J_2\!\!-\!I_i}
(\!-\!1\!)^{J_4\,\!\!+\!J_1\!\!-\!J} (\!-1\!)^{j_2\!\!+\!J_1\!\!
+\!J_2\!\!+\!J}\!\!
\sqrt{\frac{n_2(2J_2\!\!+\!1)(2J\!\!+\!1)}{(n_1\!\!
+\!n_2)}}\times
\nonumber
\\
&&\times\Big\{{J_4\,j_2\,J_2 \atop I_i\,J_1\,J}\Big\}\!\Big[j_2^{n_2-1}
(s_4\alpha_4J_4)j_2J_2|\} j_2^{n_2} (s_2\alpha_2J_2)\Big]
\left|\left(j_1^{n_1}(s_1\alpha_1J_1),j_2^{n_2-1}(s_4\alpha_4J_4)
\right)_aJ,j_2;I_i\right\rangle,   \nonumber
\end{eqnarray}
\newpage
while
\begin{eqnarray}
&& \hspace*{-0.5cm} |f\rangle =
\left| j^{n_1-1}_1\,(s'_1\alpha'_1J'_1),\,j^{n_2+1}_2\,
(s'_2\alpha'_2J'_2);I_f\,\right\rangle_a\ =
\\
&&=\sum_{s_3\alpha_3J_3J'}\!(-1)^{n_2+1+j_1\!+\!J'_2\!+\!J'_1\!+\!J'}
\sqrt{\frac{(n_1-1)(2J'_1+1)(2J'+1)}{(n_1+n_2)}} \Big\{ {J_3\,j_1\,J'_1 \atop
I_f\,J'_2\,J'} \Big\}\ \times
\nonumber
\\
&&\times\Big[j_1^{n_1\!-\!2}(s_3\alpha_3J_3)j_1J'_1|\}
j_1^{n_1-1}(s'_1\alpha'_1J'_1)\Big] \left|
\left(j_1^{n_1\!-\!2}(s_3\alpha_3J_3),
j_2^{n_2+1}(s'_2\alpha'_2J'_2)\right)_aJ',j_1;I_f\right\rangle + \nonumber
\\
&&+ \!\!\!\!\!\sum_{s_4\alpha_4J_4J}\!\!\!(\!-\!1\!)^{J'_1\!\!
+\!J'_2\!\!-\!I_f}
(\!-\!1\!)^{J_4\,\!\!+\!J'_1\!\!-\!J} (\!-1\!)^{j_2\!\!+\!J'_1\!\!
+\!J'_2\!\!+\!J}\!\!
\sqrt{\frac{(n_2\!\!+\!1)(2J'_2\!\!+\!1)(2J\!\!+\!1)}{(n_1\!\!
+\!n_2)}}\times \nonumber
\\
&&\times\Big\{{J_4\,j_2\,J'_2 \atop I_f\,J'_1\,J}\Big\}\!\Big[j_2^{n_2}
(s_4\alpha_4J_4)j_2J'_2|\} j_2^{n_2\!+\!1} (s'_2\alpha'_2J'_2)\Big]
\left|\left(j_1^{n_1\!-\!1}(s'_1\alpha'_1J'_1),j_2^{n_2}(s_4\alpha_4J_4)
\right)_aJ,j_2;I_f\right\rangle .  \nonumber
\end{eqnarray}
Here, $\Big[j^{n-1}(s'\alpha'J')jJ|\}j^{n}s\alpha J\Big]$ are the single-particle
fractional parentage coefficients (fpc) of the configuration $|j^{n}>$.
We define the reduced matrix element by the relation
\begin{eqnarray}
\langle\,I_fM_f|\hat m(\lambda\mu)|I_iM_i\,\rangle\ &=&\ (-1)^{I_f-M_f}
\Big({I_f\ \lambda\ I_i \atop -M_f\,\mu\,M_i}\Big)
\langle\,I_f||\hat m(\lambda)||I_i\rangle\,, \nonumber
\\
\langle\,I_f||\hat m(\lambda)||I_i\rangle\,&=&\,(-1)^{I_i-I_f}
\langle\,I_i||\hat m(\lambda)||I_f\rangle\,,
\end{eqnarray}
while the reduced transition rate is
\begin{equation}
B(\lambda;I_i\to I_f)\ =\
\frac{\langle I_f|| \hat m(\lambda)||I_i\rangle^2}{2I_{i}+1}\,.
\end{equation}

Then, by using formulas (2 -- 5) and the Racah algebra, we
obtain the formula for the $\beta^{+}$-decay reduced transition rate of
the multipolarity $\lambda$\, \cite{Isakov14}\,:
\begin{eqnarray}
&& \hspace*{-0.6cm}
B(\lambda;I_i\to I_f)=n_1(n_2\!+\!1)(2J_1\!+\!1)(2J'_2\!+\!1)
(2I_f\!+\!1)(2j_1\!+\!1)\
\nonumber \times
\\
&& \hspace*{-0.4cm}
\times\Big[\!j^{n_1\!-\!1}_1(s'_1\alpha'_1J'_1)j_1J_1|\}
j_1^{n_1}(s_1\alpha_1J_1)\!\Big]^2\cdot\!\Big[\!j_2^{n_2}
(s_2\alpha_2J_2)j_2J'_2
|\} j_2^{n_2\!+\!1} (s'_2\alpha'_2J'_2) \!\Big]^2
\nonumber \times
\\
&& \hspace*{-0.2cm} \times\
\left\{\begin{array}{ccc} J_1 & J_2 & I_i\\ J'_1 & J'_2 & I_f\\ j_1 &
j_2 & \lambda \end{array}\right\}^{2} B_{\rm sp} (\lambda; j_1\to j_2)\,.
\end{eqnarray}

Here, $B_{\rm sp}(\lambda; j_1 \to j_2)$ is the reduced transition rate for
the single-particle  transition.
Formula (6) is applicable for any values of the entering parameters, but
it is not visual, as in a common case the coefficients of fractional
parentage are defined from the numerical calculations (or borrowed
from the tables, see  \cite{Bayman}). However, for some simple cases
it is possible to represent the entering fractional parentage coefficients
in an obvious algebraic form. In this way, we find from \cite{Talmi} that
\begin{equation}
\Big[j^{n-1}(s_1=1, J_1=j)j\ J=0\,|\}\,j^n(s=0, J=0)\Big]\ =\ 1; \, n\mbox{ is even}
\end{equation}
and
\begin{eqnarray}
&& \hspace*{-0.7cm}
\Big[\!j^{n-1}(s_1,\,J_1\mbox{ even})j\,J\!=\!j|\}
j^n(s\!=\!1\,J\!=\!j)\!\Big]\ =
\nonumber
\\
&&=\ \left\{\!\! \begin{array}{ll}
\sqrt{\frac{(2j+2-n)}{n(2j+1)}} & \mbox{for }
s_1\!=0\ J_1\!=0 \\
-\sqrt{\frac{2(n\!-\!1)(2J_1\!+\!1)}{n(2j\!-\!1)(2j\!+\!1)}} &
\mbox{for } s_1\!=2\ J_1\neq0 ;\, n\,\, \mbox {is odd}.   \end{array} \right .
\end{eqnarray}

On the other hand, it follows from \cite{Bayman} that
\begin{eqnarray}
&& \hspace*{-0.7cm}
\Big[ j^{2j+1-n}(s_1\alpha_1J_1)j\,J\,|\}\,j^{2j+2-n}(s\,\alpha\,J)
\Big]\ =
\\
&&\hspace*{-0.2cm}=\ (\!-1)^{J_1\!+j\!-J}
\sqrt{\frac n{2j\!+\!2\!-\!n}}
\sqrt{\frac{2J_1\!\!+\!\!1}{2J\!\!+\!\!1}}
\Big[j^{n\!-1}(s\alpha J)j J_1|\} j^n(s_1\alpha_1J_1)\Big]\,.
\nonumber
\end{eqnarray}

From (8) and (9) we find that
\begin{equation}
\Big[j^{n\!-1}(s\!\!=\!1\,J\!\!=\!j)j\,J_1|\}\,j^n(s\!=\!2\,,
J_1\neq0 \mbox{ even})\Big]\!\!
=\!-\sqrt{\frac{2(2j\!+\!1\!\!-\!n)}{n(2j-1)}};\, n\,\,\mbox {is even}.
\end{equation}

From formulas  (6 -- 10), we obtain the expressions for the
transition rates in cases when the entering values of seniorities
are not more than two (note that additional quantum numbers $``\alpha"$ are
absent here). These cases are of the most interest as the corresponding
nuclear states have the lower energies.

${\bf 1.}$\,\,\, At first, we consider transitions, when both $n_1$ and
$n_2$ are even. Here arise options:

For $J_1=J_2=s_1=s_2=I_i=0$, and $J'_1=j_1,J'_2=j_2$, $s'_1=s'_2=1,
I_f=\lambda$, we have
\begin{equation}
B(\lambda; I_i=0\to I_f=\lambda)=\frac{n_1(2j_2\!+1\!-n_2)}{(2j_2+1)}\,
B_{\rm sp}(\lambda;j_1\to j_2),
\end{equation}

For $J_1=s_1=0$, $J_2\neq0$ (even), $s_2=2$, $I_i=J_2$, and $J'_1=j_1$,
$J'_2=j_2$, $s'_1=s'_2=1,$ we have
\begin{equation}
B(\lambda; I_i\!=\!J_2\!\to\!I_f)=\frac{2n_1n_2
(2I_f\!+\!1)}{(2j_2\!-\!1)} \left\{\!\!\!
\begin{array}{ccc} j_2&I_f&j_1\\ \lambda&j_2&J_2
\end{array}\!\!\!\right\}^2
B_{\rm sp}(\lambda;j_1\!\!\to\!j_2).
\end{equation}

If $J_1\neq0$ (even), $s_1=2$, $J_2=s_2=0$, $I_i=J_1$, and $J'_1=j_1$,
$J'_2=j_2$, $s'_1=s'_2=1$,  we obtain
\begin{eqnarray}
&& \hspace*{-0.5cm}
B(\lambda;I_i\!=\!J_1\!\to\!I_f)\ =\
\frac{2(2j_1\!+\!1\!-\!n_1)(2j_2\!+\!1
\!-\!n_2)}{(2j_1-1)(2j_2+1)}\times
\nonumber\\
&& \times\ (2j_1\!+\!1)(2I_f\!+\!1) \Big\{
{j_1\,I_f\,j_2 \atop \lambda\,j_1\,J_1} \Big\}^2
B_{\rm sp}(\lambda;j_1\!\to\!  j_2).
\end{eqnarray}

Finally, if both proton and neutron groups in the initial state have
seniorities two, we have ($J_1\neq0$ (even), $s_1=2$, $J_2\neq0$ (even),
$s_2=2$, and $J'_1=j_1$, $J'_2=j_2$, $s'_1=s'_2=1$)
\newpage
\begin{eqnarray}
B(\lambda;I_i\!\!\to\!\! I_f) &=&
\frac{4(2j_1\!\!+\!1\!\!-\!\!n_1)n_2}{(2j_1\!\!-\!1)(2j_2\!\!-\!\!1)}
(2J_1\!\!+\!1)(2J_2\!\!+\!\!1)(2j_1\!\!+\!\!1)(2I_f\!\!+\!\!1)
\nonumber \times
\\
&&\times \left\{ \begin{array}{ccc}
J_1 & J_2 & I_i\\ j_1 & j_2& I_f\\ j_1 & j_2 & \lambda \end{array}
\right\}^2 B_{\rm sp} (\lambda; j_1\to j_2)\,.
\end{eqnarray}

${\bf 2.}$ \,\,\, Now, consider the case when $n_1$ is even, while $n_2$
is odd. Here, we have the following options:

For $J_1=s_1=0$, $J_2=j_2$, $s_2=1$, $I_i=j_2$, and $J'_1=j_1$,
$s'_1=1$, $J'_2=s'_2=0$, $I_f=j_1$,  we have
\begin{equation}
B(\lambda; I_i=j_2\to I_f=j_1)\ =\ \frac{n_1(n_2+1)}{(2j_2+1)^2}\,
B_{\rm sp}(\lambda; j_1\to j_2).
\end{equation}

If $J_1=s_1=0$, $J_2=j_2$, $s_2=1$, $I_i=j_2$, and $J'_1=j_1, s'_1=1$,
$J'_2\neq0$ (even), $s'_2=2$,  then we obtain
\begin{eqnarray}
&& \hspace*{-0.7cm}
B(\lambda;I_i=j_2\to I_f)\,=\, \frac{2n_1(2j_2-n_2)}{(2j_2-1)(2j_2+1)}
\,(2J'_2+1)(2I_f+1)\ \times
\nonumber\\
&& \times\ \left\{ {J'_2\,I_f\,j_1\atop \lambda\ j_2\,j_2} \right\}^2
B_{\rm sp} (\lambda;j_1\to j_2)\,.
\end{eqnarray}

For $J_1\neq0$ (even), $s_1=2$, $J_2=j_2$, $s_2=1$,  and
$J'_1=j_1$, $s'_1=1$, $J'_2=s'_2=0$, $I_f=j_1$, we have
\begin{eqnarray}
&& \hspace*{-0.7cm}
B(\lambda; I_i\to I_f=j_1)\,=\,
\frac{2(2j_1\!+\!1\!-\!n_1)(n_2\!+\!1)}{(2j_1-1)(2j_2+1)}\,
(2J_1\!+\!1)(2j_1\!+\!1)\ \times
\nonumber\\
&& \times\ \left\{ {J_1\,I_i\,j_2 \atop \lambda\ j_1\,j_1} \right\}^2
B_{\rm sp}(\lambda;j_1\to j_2)\,.
\end{eqnarray}

At last, for $J_1\neq0$ (even), $s_1=2$, $J_2=j_2$, $s_2=1$,  and
$J'_1=j_1$, $s'_1=1$, $J'_2\neq0$ (even), $s'_2=2$, there follows
the result
\begin{eqnarray}
&& \hspace*{-0.7cm}
B(\lambda;I_i\!\!\to\!I_f)\!=\!\frac{4(2j_1\!\!+\!1\!\!
-\!n_1)(2j_2\!\!-\!n_2)}{(2j_1-1)(2j_2-1)}(2J_1\!\!+\!1)
(2J'_2\!\!+\!1)(2j_1\!\!+\!1)(2I_f\!\!+\!1)
\times \nonumber\\
&& \times\ \left\{ \begin{array}{ccc}
J_1 & j_2 & I_i\\
j_1 & J'_2 & I_f\\
j_1 & j_2 & \lambda \end{array} \right\}^2
B_{\rm sp} (\lambda;j_1\to j_2)\,.
\end{eqnarray}

${\bf 3.}$\,\,\, Consider now cases when $n_1$ is odd, while $n_2$ is even.
\newpage
If $J_1=j_1$, $s_1=1$, $J_2=s_2=0$, $I_i=j_1$, and $J'_1=s'_1=0$, $J'_2=j_2$,
$s'_2=1$, $I_f=j_2$, then
\begin{equation}
B(\lambda;I_i\!\!=j_1\!\!\to I_f\!=j_2)=\frac{(2j_1\!\!+\!2\!-\!n_1)
(2j_2\!+\!1\!-\!n_2)}{
(2j_1\!+\!1)(2j_2\!+\!1)} B_{\rm sp}(\lambda;j_1\!\!\to\!j_2).
\end{equation}

For $J_1=j_1$, $s_1=1$, $J_2=s_2=0$, $I_i=j_1$, and $J'_1\neq0$ (even),
$s'_1=2$, $J'_2=j_2$, $s'_2=1$,  we obtain
\begin{eqnarray}
&& \hspace*{-0.7cm}
B(\lambda;I_i\!=j_1\!\to I_f)\, =\,\frac{2(n_1\!-1)(2j_2\!+1\!-n_2)
(2J'_1\!+1)(2I_f\!+1)}{(2j_1-1)(2j_2+1)} \times
\nonumber\\
&& \times\ \left\{ {J'_1\,I_f\,j_2 \atop \lambda\ j_1\,j_1} \right\}^2
B_{\rm sp}(\lambda; j_1\to j_2)\,.
\end{eqnarray}

For $J_1=j_1$, $s_1=1$, $J_2\neq0$ (even), $s_2=2$,  and $J'_1=0$,
$s'_1=0$, $J'_2=j_2$, $s'_2=1$, $I_f=j_2$, we have
\begin{eqnarray}
&& \hspace*{-0.7cm}
B(\lambda;I_i\to I_f=j_2)\ =\ \frac{2(2j_1+2-n_1)n_2
(2J_2+1)}{(2j_2-1)}\ \times
\nonumber\\
&& \times\ \left\{ {J_2\,I_i\,j_1 \atop \lambda\ j_2\,j_2} \right\}^2
B_{\rm sp}(\lambda; j_1\to j_2)\,.
\end{eqnarray}

If $J_1=j_1$, $s_1=1$, $J_2\neq0$ (even), $s_2=2$, and
$J'_1\neq0$ (even), $s'_1=2$, $J'_2=j_2$, $s'_2=1$, we obtain
\begin{eqnarray}
&& \hspace*{-0.7cm}
B(\lambda;I_i\!\to I_f)=\frac{4(n_1\!-1)n_2(2J'_1\!+1)(2J_2\!+1)
(2I_f\!+1)(2j_1\!+1)}{(2j_1-1)(2j_2-1)} \times
\nonumber\\
&&\times\ \left\{ \begin{array}{ccc}
j_1 & J_2 & I_i\\ J'_1&j_2& I_f\\ j_1 &j_2 & \lambda \end{array}
\right\}^2 B_{\rm sp} (\lambda;j_i\to j_f)\,.
\end{eqnarray}

${\bf 4.}$\,\,\, After all, we consider cases when both $n_1$
and $n_2$ are odd.

For $J_1=j_1$, $s_1=1$, $J_2=j_2$, $s_2=1$, $I_i=\lambda$, and
$J'_1=J'_2=s'_1=s'_2=I_f=0$, we have
\begin{equation}
B(\lambda;I_i\!=\lambda\!\to I_f=0) =\frac1{(2\lambda\!+\!1)}
\frac{(n_2\!+\!1)(2j_1\!+\!2\!-\!n_1)}{(2j_2+1)}
B_{\rm sp}(\lambda;j_1\!\to j_2).
\end{equation}

If $J_1=j_1$, $s_1=1$, $J_2=j_2$, $s_2=1$, and $J'_1\neq0$ (even),
$s'_1=2$, $J'_2=0$, $s'_2=0$, $I_f=J'_1$, then
\begin{eqnarray}
&& \hspace*{-0.7cm}
B(\lambda; I_i\to I_f=J'_1)\ =\ \frac{2(n_1-1)(n_2+1)
(2J'_1+1)(2j_1+1)}{(2j_1-1)(2j_2+1)}\ \times
\nonumber\\
&& \times\ \left\{ {j_1\,j_2\,I_i \atop \lambda\ J'_1\, j_1} \right\}^2
B_{\rm sp} (\lambda;j_1\to j_2).
\end{eqnarray}

For $J_1=j_1$, $s_1=1$, $J_2=j_2$, $s_2=1$,  and $J'_1=0$,
$s'_1=0$, $J'_2\neq0$ (even), $s'_2=2$, $I_f=J'_2$, we have
\newpage
\begin{eqnarray}
&&\hspace*{-0.7cm}
B(\lambda;I_i\to I_f=J'_2)\ =\
\frac{2(2j_1+2-n_1)(2j_2-n_2)(2J'_2+1)}{(2j_2-1)}\ \times
\nonumber\\
&& \times\ \left\{ {j_2\,j_1\,I_i \atop \lambda\ J'_2\,j_2} \right\}^2
B_{\rm sp}(\lambda;j_1\to j_2)\,.
\end{eqnarray}

In the case of $J_1=j_1$, $s_1=1$, $J_2=j_2$, $s_2=1$,  and
$J'_1\neq0$ (even), $s'_1=2$, $J'_2\neq0$ (even), $s'_2=2$, we
obtain the result
\begin{eqnarray}
&& \hspace*{-0.7cm}
B(\lambda;I_i\!\!\to\! I_f)=\frac{4(n_1\!\!-\!1)(2j_2\!\!-\!n_2)
(2J'_1\!\!+\!1)
(2J'_2\!\!+\!1)(2I_f\!\!+\!1)(2j_1\!\!+\!1)}{(2j_1-1)(2j_2-1)}
\nonumber \times
\\
&& \times\ \left\{ \begin{array}{ccc}
j_1 & j_2 & I_i\\
J'_1 & J'_2 & I_f\\
j_1 & j_2 & \lambda \end{array} \right\}^2 B_{\rm sp} (\lambda;j_1\to j_2).
\end{eqnarray}

The aforesaid formulas are suitable for any beta-transitions.
One should mention here that different from \cite{Isakov09},
the foregoing calculations are
performed in the neutron--proton representation, where the wave
functions do not have definite isospin. This is most important for
Fermi transitions, where the transition operator is nothing but the
rotation in the isospin space,
$\hat{T}_x \pm i\, \hat{T}_y = \mp\sqrt2\,\, \hat{T}_{\pm 1}$.
Here, transition
proceeds only between the isoanalogue states with the  probability
$B(F)=T(T+1)-T_{z_{i}}T_{z_{f}}$.  Among the other transitions
the Gamow--Teller ones are of the most importance. Thus, we quote
necessary formulas for calculation of the corresponding single-particle
reduced matrix elements $j^{\pm}=l\pm1/2$:
\begin{eqnarray}
&& \langle n\ell j^+\|\hat m(\lambda=1,GT)\|n\ell j^+\rangle\ =\
\sqrt{\frac{(2\ell+2)(2\ell+3)}{2\ell+1} }\,,\nonumber
\\
&&\langle n\ell j^-\|\hat m (\lambda=1,GT)\|n\ell j^+\rangle\ =\
\sqrt{\frac{2(2\ell)(2\ell+2)}{2\ell+1}}\,,
\\
&& \langle n\ell j^-\| \hat m(\lambda=1, GT)\|n\ell j^-\rangle\ =\
 -\sqrt{\frac{(2\ell-1)(2\ell)}{2\ell+1} }\,. \nonumber
\end{eqnarray}

Note that the formulas shown by us, are of course
applicable for the $\beta^-$\,-decay. In this case $\{ j_1 \}$ refers to
neutrons, while $\{ j_2 \}$ -- to protons.
The formulas may also be used for description of $\gamma$\,-decay.
In this case both orbitals refer to the same sort of nucleons, {\it while the
transition  happens between the nucleons of different groups}.

Consider now gamma- transitions within the configuration
$\Big\{j_1^{n_1},\,j_2^{n_2}\Big\},$\, i.e.
when the initial and final states have the form
\begin{eqnarray} 
&& |i\rangle\ =\ |j_1^{n_1}(s_1\alpha_1J_1)\,j^{n_2}_2(s_2\alpha_2J_2);
I_i\rangle_a\,\,,\,\,{\rm and} \nonumber
\\
&& |f\rangle\ =\ |j^{n_1}_1 (s'_1\alpha'_1J'_1), j^{n_2}_2
(s'_2\alpha'_2J'_2); I_f\rangle_a\,.
\end{eqnarray}

Again, by using the decomposition (2) and Racah technique, we
obtain the expression for the transition matrix element between the
states $|j^{n_1}_1(s_1\alpha_1J_1),
j_2^{n_2}(s_2\alpha_2J_2);I_i\rangle$ and
$|j_1^{n_1}(s'_1\alpha'_1J'_1),j_2^{n_2}
(s'_2\alpha'_2J'_2);I_f\rangle$ \, \cite{Isakov15}\,:

\begin{eqnarray} 
&& \hspace*{-0.8cm}
\langle I_f\|\sum^{n_1+n_2}_{k=1} m_\lambda(k)\| I_i\rangle\ =\ n_1
\sqrt{(2J_1+1)(2J'_1+1)(2I_i+1)(2I_f+1)}\ \langle
j_1\|m_\lambda\|j_1\rangle\ \times
\nonumber\\
&&\times\ \delta(s_2\alpha_2J_2,s'_2\alpha'_2J'_2)
\left\{{J_2\,I_i\,J_1\atop \lambda\,J'_1\,I_f}\right\}
\sum_{s\alpha J}\  (-1)^{J_1+J'_1+J_2+j_1+I_i+J}\ \times
\nonumber\\
&&\times\ \Big[j_1^{n_1-1}(s\alpha J)j_1J_1|\!\}
j_1^{n_1}(s_1\alpha_1J_1) \Big]  \Big[ j_1^{n_1-1}(s\alpha J)
j_1J'_1|\!\} j_1^{n_1} (s'_1\alpha'_1J'_1)\Big]
\left\{{J\,j_1\,J_1 \atop \lambda\,J'_1j_1}\right\}\ +
\nonumber\\
&& +\ n_2\sqrt{(2J_2+1)(2J'_2+1)(2I_i+1)(2I_f+1)}\
\langle j_2\|m_\lambda\|j_2\rangle\
\delta(s_1\alpha_1J_1,s'_1\alpha'_1J'_1)\ \times
\nonumber\\
&& \times\ \left\{{J_1I_iJ_2\atop \lambda\,J'_2I_f}\right\}
\sum_{s\alpha J}\, (-1)^{J_2+J'_2+J_1+j_2+I_f+J}
\Big[j_2^{n_2-1}(s\alpha J)j_2J_2|\!\}\,j_2^{n_2}(s_2\alpha_2J_2)\Big]
\times
\nonumber\\
&&\times\ \Big[j_2^{n_2-1}(s\alpha J)j_2J'_2|\!\} j_2^{n_2}
(s'_2\alpha'_2J'_2)\Big]
\left\{{J\,j_2\,J_2\atop \lambda\,J'_2\,j_2}\right\}.
\end{eqnarray}

One can simplify formula (29) if the transition proceeds between the
states\\
$|i\rangle=|j_1^{n_1}(s_1=1,J_1=j_1), j_2^{n_2}(s_2=1,
J_2=j_2);I_i\rangle$ and \\ $|f\rangle=|j_1^{n_1}(s'_1=1,J'_1=j_1),
j_2^{n_2}(s'_2=1, J'_2=j_2); I_f\rangle$,
if we use algebraic form (8) for the entering coefficients
of fractional parentage.
Then we have simple formula for the transition matrix element:
\begin{eqnarray}  
&& \hspace*{-0.8cm}
\langle I_f\| \sum^{n_1+n_2}_{k=1}\ m_\lambda(k)\| I_i\rangle\ =\
\sqrt{(2I_i+1)(2I_f+1)} \Bigg[ (-1)^{j_1+j_2+I_i+\lambda}
\left\{ {j_2I_ij_1\atop \lambda\,j_1I_f}\right\} \times
\nonumber\\
&&\times\ \langle j_1\| m_\lambda\| j_1\rangle
\frac{2j_1-n_1+(-1)^{\lambda+1}(n_1-1)}{2j_1-1}
+(-1)^{j_1+j_2+I_f+\lambda}\ \times
\nonumber\\
&& \times\ \left\{{j_1 I_f j_2\atop \lambda\,j_2I_i} \right\}
\langle j_2\|m_\lambda\|j_2\rangle\,
\frac{2j_2-n_2+(-1)^{\lambda+1}(n_2-1)}{2j_2-1}\Bigg].
\end{eqnarray}

Mention that for isolated $j$-level and accounting for the
blocking effect \cite{Soloviev}, we have the expression for the
$u,v$- coefficients of the Bogoliubov transformation:
\begin{equation}  
u^2_j\ =\ \frac{2j-n_j}{2j-1}\,; \quad v^2_j\ =\ \frac{n_j-1}{2j-1}\,.
\end{equation}
Then, in formula (30) we can introduce the notation
\begin{equation} 
\frac{2j-n+(-1)^{\lambda+1}(n-1)}{2j-1}\ = \left\{ \begin{array}{cccl}
\displaystyle\frac{2j-2n+1}{2j-1} &\to & u^2_j-v^2_j & \mbox{for
$E\lambda$ operators},\\
1 & \to & u^2_j+v^2_j=1 & \mbox{for $M\lambda$ operators.}\end{array}
\right.
\end{equation}
In this form formula (30) remains also valid, when the levels $\{j\}$
are not isolated ones.

From formula (30) one can easily obtain expressions for the electric
quadrupole and magnetic dipole moments of the state
$|j_1^{n_1}(s_1=1,J_1=j_1),j_2^{n_2}(s_2=1,J_2=j_2);I\rangle$. In this
case, we have
\begin{eqnarray} 
O_2(n_1j_1,n_2j_2;I)&=&\sqrt{\frac{I(2I-1)(2I+1)}{(I+1)(2I+3)}}
\Bigg[\left\{{j_2j_1I\atop 2\,I\,j_1}\right\}
\sqrt{\frac{(2j_1+1)(j_1+1)(2j_1+3)}{j_1(2j_1-1)}}q_2(n_1,j_1)+ \nonumber
\\
&+&\left\{{j_1j_2I\atop
2\,I\,j_2}\right\} \sqrt{\frac{(2j_2\!+\!1)(j_2\!+\!1)(2j_2\!+\!3)}{j_2(2j_2-1)}}
\,q_2(n_2,j_2)\Bigg](-1)^{j_1+j_2+I}.
\end{eqnarray}
Here, $q_2(n,j)$ is the quadrupole moment of $n$ particles having seniority
$s=1$ on the isolated $j$-level:
\begin{equation}  
q_2(n,j)\ =\ - e_{eff}\cdot\frac{(2j-2n+1)}{(2j+2)}\cdot
\langle  j|r^2|j\rangle\,.
\end{equation}
In the case of non-isolated level one should also make in (34) the
substitution (32).

For magnetic moments formula (30) strongly simplifies. As a result, we
have the expression for a gyromagnetic ratio that does not depend on
filling of subshells:
\begin{equation} 
g_I\ =\ \frac{g_1+g_2}2 +\frac{g_1-g_2}2\cdot
\frac{j_1(j_1+1)-j_2(j_2+1)}{I(I+1)}\,.
\end{equation}

Another case, when one can simplify formula (29) arises when the
transition proceeds between the states $|i\rangle = |j_1^{n_1}(s_1=2,
J_1\neq0, {\rm even}), j_2^{n_2}(s_2=0,J_2=0); I_i=J_1\rangle$ and
$|f\rangle = |j_1^{n_1}(s_1^{'}=0, J_1^{'}=0),
j_2^{n_2}(s_2^{'}=0,J_2^{'}=0); I_f=0\rangle$, when both $n_1$ and
$n_2$ are even numbers. Here, by using formulas (7) and (10)
we obtain for $E\lambda$-transitions the result ; $\lambda$ is even:
\begin{equation}  
B(E\lambda; I_{i} \to I_{f}) =
\frac{2n_1(2j_1+1-n_1)}{(2j_1-1)(2j_1+1)(2\lambda+1)}\,
\langle j_1\| m_\lambda\| j_1\rangle^2\,\, \delta(I_i, J_1)\,\,
\delta(J_1,\lambda)\,\,\delta(I_f,0)\,.
\end{equation}
The subshell $\{j_2\}$ is inert here.

In light and medium-$A$ nuclei, which have $N\sim Z$, we reveal decays when
$n_1=n_2=n$, $\ell_1=\ell_2=\ell$, and $j_1=j_2=j$. Here, we may use the
single-group isospin representation and may characterize wave functions of
both initial and final states in the form, see \cite{Talmi}\,:
\begin{eqnarray}
&& |i\rangle\ =\ |j^{n_t}(s_i\,\alpha_i\,t_i\,\beta_i);\,J_i\,M_i,T_i\,
T_{z_i}\rangle_a\ \mbox{ and}
\nonumber\\
&& |f\rangle\ =\ |j^{n_t}(s_f\alpha_f t_f \beta_f);\,
J_fM_f,T_fT_{z_f}\rangle_a\, ,
\end{eqnarray}
where $n_t$ is total number of protons and neutrons on the $j$-orbit,
$t$ is the reduced isospin (isospin of unpaired nucleons), while
$\beta$ is an additional quantum number in the isospin space, analogous
to $\alpha$ in the spin--coordinate space. In this case, the reduced
(over the angular momentum) transition rate for the $\beta^{\pm}$
decays of the multipolarity $\lambda$ reads as
\begin{eqnarray}
&& \hspace*{-0.5cm}
B^{\pm}(\lambda;J_i\to J_f)\,=\,\delta(T_{z_f},T_{z_i}\pm1)\Bigg[
\sum_{TJs\alpha t\beta}(-1)^{J+j+J_f+\lambda+T_i+T+\frac12}\sqrt{2J_f+1}
\Big\{{j\,J_f\,J \atop \!\!J_i\,j\,\,\lambda}\Big\} \times
\nonumber\\
&&\times\ \langle j\|\hat m(\lambda)\| j\rangle\,n_t\,\Big[j^{n_t-1}
(s\alpha t\beta JT) jJ_iT_i|\} j^{n_t}(s_i\,\alpha_i\,t_i\,\beta_i\,
J_i\,T_i)\Big] \times
\\
&&\hspace{-0.5cm}\times\ \Big[ j^{n_t-1}(s\alpha t\beta JT)jJ_fT_f|\}
j^{n_t} (s_f\alpha_ft_f\beta_fJ_fT_f)\Big]\,\sqrt{3(2T_f+1)}\,
C^{T_iT_{z_i}}_{T_fT_{z_i}\pm 1\,\,1\mp 1} \Big\{{T_iT_f1\atop\frac12\,\frac12\,T}\Big\}\Bigg]^2.
\nonumber
\end{eqnarray}
Here, $\Big[ j^{n-1}(s'\alpha' t'\beta' J'T')jJT|\}j^{n} (s\alpha t\beta JT)\Big]$
are fractional parentage coefficients in the
isospin representation. For $j \le 7/2$ these coefficients are tabulated in
\cite{Towner69,Towner97}.

In case of beta-decay, the transition operator is of the isovector
nature, and is proportional to $\frac{1}{\sqrt{2}}\hat\tau_{\pm1}$.
The situation is
different in case of electromagnetic transitions. As the
electromagnetic field differently interacts with protons and neutrons,
the corresponding transition operator in the isospin representation
contains both isovector and isoscalar contributions which are both diagonal
over the third projection of isospin, notably
\begin{eqnarray}
\hat m(\lambda)\ &=& \ \hat m(\lambda,T=0)+\hat m(\lambda,T=1)\cdot\hat\tau_0\,,
\,\,\, {\rm where}
\\
\hat m(\lambda,T=0) &=& \frac{\hat m(\lambda,n)+\hat m(\lambda,p)}2\,,
\quad \hat m(\lambda,T=1)=\frac{\hat m(\lambda,n)-\hat
m(\lambda,p)}2\,.
\nonumber
\end{eqnarray}
After some calculations, we obtain the result:
\begin{eqnarray}
&& \hspace*{-0.5cm}
B^{\rm em}(\lambda;\,J_i\to J_f)\ =\ \delta(T_{z_f},T_{z_i}) \Bigg[
\sum_{TJs\alpha t\beta} (-1)^{J+j+J_f+\lambda} \sqrt{2J_f+1}\,\, n_t\
\times \nonumber
\\
&& \times\Big\{{j\,J_fJ \atop J_i\, j\,\lambda}\Big\} \Big[j^{n_t-1}(s\alpha
t\beta JT)jJ_i\,T_i |\}\, j^{n_t}(s_i\,\alpha_i\,t_i\,\beta_i\,J_i\,T_i)\Big]
\times \\
&&\times \Big[j^{n_t-1}(s\alpha t\beta JT)j J_fT_f|\} j^{n_t}
(s_f\alpha_ft_f\beta_fJ_fT_f)\Big] \bigg( \delta(T_f,T_i) \langle j\|
\hat m(\lambda, \,0)\|j \rangle +  \nonumber
\\
&&+\ (-1)^{T_i+T-\frac12} \sqrt{6(2T_f+1)}\,\langle j\|\hat m
(\lambda, \,1)\|j\rangle\, C^{T_iT_{z_i}}_{T_fT_{z_i}\,1\,0}
\Big\{ {T_iT_f1 \atop \frac12\,\frac12\,T}\Big\} \bigg)\Bigg]^2.
\nonumber
\end{eqnarray}

For ease, we below show formulas for the reduced matrix elements of the
single-particle electromagnetic transition operators. As compared to
\cite{Bohr69}, the formulas are reduced to the most simple form:
\begin{eqnarray} && \hspace*{-1.2cm}
\langle n_1\ell_1j_1\| \hat
m(E\lambda)\|n_2\ell_2j_2\rangle\ =\
e_{\rm eff}\langle1|r^\lambda|2\rangle(-1)^{\lambda+j_1-1/2}\ \times
\\
&& \times\ \Big[\frac{(2j_1+1)(2j_2+1)}{4\pi}\Big]^{1/2}
\frac{[1+(-1)^{\ell_1+\ell_2+\lambda}]}2 C^{\lambda\,\,0}_{j_1\,
\frac12\,j_2-\frac12}\,\,,
\nonumber
\end{eqnarray}

\begin{eqnarray}
&& \hspace*{-0.5cm}
\langle n_1\ell_1j_1\| \hat m(M\lambda)\|n_2\ell_2j_2\rangle\ =
\\
&&=\ \frac{e\hbar}{2m_Nc}\langle1|r^{\lambda\!-1}|2\rangle
(-1)^{\lambda\!+j_1\!+\frac12}
\Big[\frac{(2j_1\!+\!1)(2j_2\!+\!1)}{4\pi}\Big]^{\frac12}
\frac{[1\!+\!(-1)^{\ell_1\!+\ell_2\!+\lambda+1}]}2 \times
\nonumber\\
&&\times\Bigg\{\Big(g_s\!-\frac2{\lambda\!+\!1}g_\ell\Big)\frac\lambda2
C^{\lambda\,\,0}_{j_1\!\frac12\,j_2\!-\frac12}
\bigg[1\!+\!\frac{(-1)^{\ell_2\!+\!j_2\!-\!\frac12}}\lambda
\Big\{\Big(j_2\!+\!\frac12\Big)\!+\!(-1)^{j_1\!+\!j_2\!+\!\lambda}
\Big(j_1\!+\!\frac12\Big)\Big\}\bigg]+
\nonumber\\
&&+ \frac{g_\ell}{\lambda\!+1}
\Big[\frac{2\lambda\!+1}{2\lambda\!-1}\Big]^{\frac12}
C^{\lambda-1\,\,\,0}_{j_1\!\frac12\,j_2-\frac12}
\Big[(j_1\!+j_2\!+\lambda\!+1)(j_1\!+j_2\!-\lambda\!+1)(j_1\!+\lambda\!-j_2)
(j_2\!+\lambda\!-j_1)\Big]^{\frac12}\Bigg\}.
\nonumber
\end{eqnarray}

Here we used phases of spherical harmonics as in \cite{Condon35},
$Y_{l-m}= (-1)^{m}\,Y_{lm}^{*}$.

For the purpose of qualitative evaluations, we also present expressions
for some radial matrix elements, obtained by using the oscillator
potential:
\begin{eqnarray}
&&  \langle n'\,\ell\!+1|r|n\ell\rangle\ =\
\Big(\frac\hbar{m\omega}\Big)^{\frac12}
\left[\delta_{n',n}\sqrt{n+\ell+3/2} -\delta_{n',n-1}\sqrt{n}\right],
\end{eqnarray}
\begin{eqnarray}
&& \hspace*{-0.7cm}
\langle n'\,\ell+2\,|r^2|\,n\ell\rangle\ =\ \frac\hbar{m\omega}\
\times  \\
&& \times \left[\!\,\delta_{n',n\!-\!2}
\sqrt{n(n\!-\!1)}\!-\!\delta_{n',n\!-\!1}
2\sqrt{n(n\!+\!\ell\!+\!3/2)}\!+\!\delta_{n',n}
\sqrt{(n\!+\!\ell\!+\!3/2)(n\!+\!\ell\!+\!5/2)}\!\,\,\right],
\nonumber
\\
&& \hspace*{-0.7cm}
\langle n'\,\ell\, |r^2|\,n\ell\rangle\ =\ \frac\hbar{m\omega}\ \times
\\
&& \times \left[\!\,\delta_{n',n} (2n+\ell+3/2)
-\delta_{n',n\!-1}\sqrt{n(n+\ell+1/2)} - \delta_{n',n\!+\!1}
\sqrt{(n+1)(n+\ell+3/2)}\right],
\nonumber
\end{eqnarray}

\begin{eqnarray}
&& \hspace*{-0.6cm}
\langle n'\,\ell+1\,|r^3| n\ell\rangle\ =\
\left(\frac\hbar{m\omega}\right)^{3/2}\ \times
\\
&& \times\  \bigg[\delta_{n',n-2} \sqrt{(n-1)n(n+\ell+1/2)}
-\delta_{n',n-1} (3n+2\ell+2)\sqrt n\ +
\nonumber\\
&&+\ \delta_{n',n}\Big(3n+\ell+5/2\Big) \sqrt{n+\ell+3/2}
-\delta_{n',n+1} \sqrt{(n+1)(n+\ell+3/2)(n+\ell+5/2)}\bigg],
\nonumber
\end{eqnarray}

\begin{eqnarray}
&& \hspace*{-0.6cm}
\langle n'\,\ell+3\,|r^3|\,n\ell\rangle\ =\
\left(\frac\hbar{m\omega}\right)^{3/2}\ \cdot
\ \bigg[ \delta_{n',n} \sqrt{(n+\ell+3/2)(n+\ell+5/2)
(n+\ell+7/2)}\ -
\nonumber
\\
&&-\ \delta_{n',n-1}\,3\sqrt{n(n+\ell+3/2)(n+\ell+5/2)}\ +
\ \delta_{n',n\!-2}\,3\sqrt{(n\!-\!1)n(n\!+\!\ell\!+\!3/2)}-
\\
&&-\delta_{n',n-3}\sqrt{(n\!-\!2)(n\!-\!1)n}\bigg].
\nonumber
\end{eqnarray}

Here, $\hbar/m\omega \approx1.01\,A^{1/3}\rm fm^2$ by
$\hbar\omega=41/A^{1/3}\,$MeV, while $n$ is the radial quantum number
 $(n=0,1,2,... $; for instance, $n=0$ for the $1s$-state); the phases
 of radial functions are defined by the condition $R_{nl}(r\to 0) > 0$.

The formulas derived by us are valid for analysis of experimental data,
especially concerning nuclei in the regions of magicity, and enable
to perform  qualitative evaluations without complicated numerical
calculations.

\end{document}